\documentclass[12pt,preprint]{emulateapj}

\usepackage{color}
\usepackage{ulem}

\slugcomment{Accepted by The Astrophysical Journal (2008, Mar)
[astro-ph version w/low-resolution images]}

\shorttitle{Star Formation Extreme Outer Galaxy}
\shortauthors{Kobayashi et al.}

\begin{document}


\title{Star Formation in the Most Distant Molecular Cloud\\ in the
Extreme Outer Galaxy:\\ A Laboratory of Star Formation in an
Early Epoch of the Galaxy's Formation\altaffilmark{1,2,3}}


\author{Naoto Kobayashi\altaffilmark{4,5}, Chikako Yasui\altaffilmark{}}
\affil{Institute of Astronomy, School of Science, University of Tokyo,
    2-21-1 Osawa, Mitaka, Tokyo 181-0015, Japan}
\email{naoto@ioa.s.u-tokyo.ac.jp}

\author{Alan T. Tokunaga\altaffilmark{4}}
\affil{Institute for Astronomy, University of Hawaii, 2680 Woodlawn
Drive, Honolulu, HI 96822}

\and

\author{Masao Saito\altaffilmark{}}
\affil{ALMA Project, National Astronomical Observatory of Japan, 2-21-1
Osawa, Mitaka, Tokyo 181-8588, Japan}


\altaffiltext{1}{ Based on data collected at University of Hawaii 2.2 m
telescope.  }

\altaffiltext{2}{ This research has made use of the NASA/IPAC Infrared
Science Archive, which is operated by the Jet Propulsion Laboratory,
California Institute of Technology, under contract with the National
Aeronautics and Space Administration.  }

\altaffiltext{3}{ This research used the facilities of the Canadian
Astronomy Data Centre operated by the National Research Council of
Canada with the support of the Canadian Space Agency.  }

\altaffiltext{4}{Visiting Astronomer, University of Hawaii 2.2 m
Telescope.
}  

\altaffiltext{5}{Half of the work was done when visiting at the
University of Hawaii during 1996--1999.}


\begin{abstract}

We report the discovery of active star formation in Digel's Cloud 2,
which is one of the most distant giant molecular clouds known in the
extreme outer Galaxy (EOG).  At the probable Galactic radius of $\sim$20
kpc, Cloud 2 has a quite different environment from that in the solar
neighborhood, including lower metallicity, much lower gas density, and
small or no perturbation from spiral arms. With new wide-field
near-infrared (NIR) imaging that covers the entire Cloud 2, we
discovered two young embedded star clusters located in the two dense
cores of the cloud. Using our NIR and $^{12}$CO data as well as
\ion{H}{1}, radio continuum, and IRAS data in the archives, we discuss
the detailed star formation processes in this unique environment.
We show clear evidences of a sequential star formation triggered by the
nearby huge supernova remnant, GSH 138-01-94.  The two embedded clusters
show a distinct morphology difference: the one in the northern molecular
cloud core is a loose association with isolated-mode star formation,
while the other in the southern molecular cloud core is a dense cluster
with cluster-mode star formation. We propose that high compression by
the combination of the SNR shell and an adjacent shell caused the dense
cluster formation in the southern core.
Along with the low metallicity range of the EOG, we suggest that EOG
could be an excellent laboratory for the study of star formation
processes, such as those triggered by supernovae, that occured during an
early epoch of the Galaxy's formation. In particular, the study of
the EOG may shed light on the origin and role of the thick disk, whose
metallicity range matches with that of the EOG well.

\end{abstract}




\section{INTRODUCTION}


The extreme outer Galaxy (EOG), which we define as the region with a
Galactic radius ($R_\mathrm{g}$) of more than 18 kpc, has a very
different environment from the regions near our solar system, since it
has a much lower gas density, lower metallicity (see the metallicity
gradient curve in, e.g., Smartt \& Rolleston 1997) and small or no
perturbation from spiral arms. Such a region is not only of strong
interest in itself, but it also provides a good opportunity to study
astronomical processes, such as star formation,
\textcolor{black}{ in a physical environment that is very different from
that of the solar neighborhood. While the detailed star formation
processes have been studied mostly for nearby star-forming regions at
the distances less than 1 kpc, where the physical/chemical environment
appears to be relatively uniform, the EOG enables us to study how the
environment (density, temperature, pressure, external radiation field,
metallicity, etc.) affects the basic star formation processes and
parameters such as the initial mass function, star formation efficiency,
and disk formation efficiency.}


Wouterloot \& Brand (1989), Wouterloot et al. (1990), Brand \&
Wouterloot (1995) obtained the first census of molecular clouds in the
outer Galaxy based on the CO survey of IRAS point sources. Digel, de
Geus, \& Thaddeus (1994) found eleven molecular clouds, including Cloud
2, in the EOG based on CO observations of distant \ion{H}{1} peaks in
the Maryland-Green Bank survey (Westerhout \& Wendlant 1982). Following
those pioneering works, Heyer et al. (1998) conducted a comprehensive CO
survey of molecular clouds in the outer Galaxy with the Five College
Radio Astronomy Observatory (FCRAO) CO survey.
\textcolor{black}{Recently, Ruffle et al. (2007) conducted a molecular
line survey of some of the Digel clouds and provided maps and chemical
modeling of Cloud 2. Such detailed study will eventually reveal how
those molecular clouds were formed in the EOG.}
As for the star formation activity, de Geus et
al. (1993) first found an \ion{H}{2} region associated with Digel's Cloud
2. Later Kobayashi \& Tokunaga (2000) found associated red infrared
sources to confirm the star-forming activity in this cloud. Santos et
al. (2000) and later Snell, Carpenter, \& Heyer (2002) found 
\textcolor{black}{a considerable number of embedded star clusters in
molecular clouds in the outer Galaxy that confirm star formation in
clusters is common in the outer Galaxy up to $R_\mathrm{g}$ $\sim$ 17
kpc}.
Recently, Brand \& Wouterloot (2007) also reported a discovery of an
embedded cluster in a molecular cloud, WB89-789, at the probable
Galactic radius of $\sim$20 kpc, which marks one of the most distant
embedded clusters in the EOG.


Here we report the discovery of two embedded clusters in Digel's Cloud
2, which is one of the most distant molecular clouds in the EOG (Digel
et al. 1994). 
\textcolor{black}{While conducting a near-infrared survey of the Digel
clouds to study the star formation activity in the EOG (Kobayashi \&
Tokunaga 2001), we found clear evidence of star formation activity in
Cloud 2 (Kobayashi \& Tokunaga 2000). As a follow-up study, we are
extensively studying this cloud with deep near-infrared imaging using
the Subaru 8.2 m telescope (Yasui et al. 2006, 2008) and with}
CO observations using the Nobeyama 45 m telescope (M. Saito et al. 2008,
in preparation).


Originally, the kinematic distance to Cloud 2 was estimated at
$R_\mathrm{g} = 28$ kpc (heliocentric distance: $D = 21$ kpc), while the
latest \ion{H}{1} observation by Stil \& Irwin (2001) suggests
$R_\mathrm{g} = 23.6$ kpc ($D = 16.6$ kpc).
\textcolor{black}{Ruffle et al. (2007) use $R_\mathrm{g}$ = 22 and 28 kpc
for their distance sensitive calculations. However, several authors
questioned this rather high $R_\mathrm{g}$ from independent distance
estimates of the associated early-type star, MR-1 (Muzzio \& Rydgren
1974). The association of MR-1 with Cloud 2 appears to be quite firm
based on the association of the H$\alpha$ nebulae (de Geus et al. 1993)
and the photodissociation region (Kobayashi \& Tokunaga 2000), and
because all three components have almost the same LSR velocity: $-102.4$
km s$^{-1}$ for Cloud 2 (Digel et al. 1994); $-102.7$ $\pm$ 12 km
s$^{-1}$ for MR-1 (Russeil, Adami, \& Georgelin 2007) and $-$101 km
s$^{-1}$ for the H$\alpha$ nebulae (de Geus et al. 1993). Smartt,
Dufton, \& Rolleston (1996) estimated the Galactic radius of MR-1 as
$R_\mathrm{g} = 15 - 19$ kpc ($D = 8 - 12$ kpc) with high-resolution
optical spectra: the shorter and the longer distances are based on LTE
and non-LTE model stellar atmospheres.  Recently, Russeil et al. (2007)
re-estimated the distance to MR-1 as $R_\mathrm{g} = 14.3 \pm 0.5$ kpc
($D = 6.78 \pm 0.59$ kpc) using a newly obtained high-resolution optical
spectrum and an LTE model stellar atmosphere, which is consistent with
the LTE distance ($D = 8$ kpc) by Smartt et al. (1996).
Because the spectroscopic distance of stars should be more accurate than
kinematic distance, the actual distance to Cloud 2 is likely to be that
of MR-1, which is less than the kinematic distance calculated from
\ion{H}{1}/CO data.
Throughout this paper, we adopt the most likely distance $R_\mathrm{g} =
19$ kpc ($D = 12$ kpc) for Cloud 2, the same one used in our previous
paper (Kobayashi \& Tokunaga 2000) because a non-LTE model is more
likely to be accurate for stars in the effective temperature regime of
MR-1 (Smartt et al. 1996). It is also the largest distance among those
estimated by stellar spectroscopy, and more consistent with the radio
kinematic distances.
When we discuss about distance sensitive parameters, we will mention the
possible systematic uncertainty.
}


The ambient \ion{H}{1} density at $R_\mathrm{g}$ $\sim$ 20 kpc is
thought to be very low because of the small surface density and the
large-scale height: $N_\mathrm{HI}$ could be as low as 0.001 cm$^{-3}$
(e.g., Nakanishi \& Sofue 2003). The metallicity of Cloud 2 is estimated
at $-0.7$ dex from the radio molecular lines (Lubowich et al. 2004; see
also Ruffle et al. 2005), which is consistent with the metallicity of
MR-1 as measured by optical spectroscopy ($-0.5$ to $-0.8$ dex: Smartt
et al. 1996; Rolleston et al. 2000).


The latest \ion{H}{1} observation by Stil \& Irwin (2001) revealed that
a remarkably large supernovae remnant (SNR) shell, GSH 138-01-94, is
associated with Cloud 2. 
Because there is little or no perturbation from the spiral arms in the
EOG enviroment,
\textcolor{black}{the supernova (SN) could have triggered star formation
in Cloud 2. It is usually hard to determine what triggers star formation
in the inner part of the Galactic disk because of many chance
projections of various foreground and background objects and because the
observed star-forming region itself has a complex structure of gas/dust
and stars from its long star formation history and high ambient gas
density. In such a region, a SNR shell cannot keep its uniform shape for
a long enough time (e.g., a few Myr) to discern the SN-triggered star
formation process because of the large amount of ambient material
distributed nonuniformly. However, the SNR shell expansion can be
clearly observed in the EOG, which is free from the above
complexity. Actually, GSH 138-01-94 shows a large complete SNR shell
that has lasted for more than 4 Myr, and is the largest and oldest SNR
known in the Galaxy (Stil \& Irwin 2001).}
Therefore, Cloud 2 is an excellent place to study SN-triggered star
formation, which is thought to be one of the major star formation
processes in galaxies (e.g., Elmegreen 2002).


The low metallicities in the EOG regions are comparable to those for
Galactic thick disk stars (e.g., Brewer \& Carney 2006), dwarf irregular
galaxies, and damped Lyman-$\alpha$ systems (e.g., Pettini 2004). In
conjunction with the very low gas density and lack of spiral arm
perturbation, the EOG may approximate the environment of star formation
in the early universe.
\textcolor{black}{ Because dwarf irregular galaxies have a similar
environment--- low-metallicity, low-gas density, and lack of spiral arm
perturbation, they are also thought to represent the conditions of star
formation in the early universe (e.g., Hunter, Elmegreen, \& Martin
2006). However, the EOG is advantageous for the study of such star
formation processes simply because of its proximity compared to the
galaxies in the local univserse. }
Studies of the chemical composition of halo stars in our Galaxy show
that star formation in the early epoch of the Galaxy's formation was
mostly triggered by supernovae (Ryan, Norris, \& Beers 1996; Audouze \&
Silk 1995; Shigeyama \& Tsujimoto 1998). Therefore, {\it Cloud 2 and
other EOG star-forming regions are excellent laboratories in which to
study the star formation process in an environment that is similar to
that which existed during an early epoch of the Galaxy's formation}.


In this paper, we discuss our near-infrared as well as $^{12}$CO
observational results for Cloud 2 in the above context. We also make use
of the mid-infrared (IRAS), \ion{H}{1}, and radio continuum data in the
archives for our discussion. The companion papers by Yasui et al. (2006,
2008) discuss very deep infrared imaging of the star-forming clusters
with the Subaru 8.2 m telescope, and they reinforce our interpretation.

\section{NEAR-INFRARED IMAGING}

Near-infrared images of Cloud 2 in the standard $J$-, $H$-, and $K$-band
filters were obtained over a five-night period (3$-$7 October 1998)
using the QUick InfRared Camera (QUIRC) at the University of Hawaii 2.2
m telescope on Mauna Kea. QUIRC uses a 1024 $\times$ 1024 HgCdTe
Astronomical Wide Area Infrared Imaging (HAWAII) array and was used at
the f/10 focus to provide a plate scale of 0$\farcs$1886 pixel$^{-1}$
with a field of view of roughly 3$\farcm$2 $\times$ 3$\farcm$2. The
entire arc-shaped Cloud 2 was covered with a mosaic of QUIRC fields.
For the present paper, we used only the dataset of the third night, when
the seeing conditions were best ($\sim$0$\farcs$5) and the star clusters
were clearly resolved. The sky was covered with occasional thin cirrus
on that night, but the photometric uncertainty due to cirrus was
insignificant. The total integration time was 5, 10, 15 min for $K$-,
$H$-, and $J$-bands, respectively.


All the data for each band were reduced with IRAF\footnote{IRAF is
distributed by the National Optical Astronomy Observatories, which are
operated by the Association of Universities for Research in Astronomy,
Inc., under cooperative agreement with the National Science Foundation.}
with standard procedures: dark subtraction, flat-fielding, bad-pixel
correction, median-sky subtraction, image shifts with dithering offsets,
and combining. The images of eight-fields were combined into one large
mosaic with standard IRAF image matching tasks.



Figure 1 shows a $JHK$ three-color image of Cloud 2 with $^{12}$CO (1-0)
contours overplotted (the radio data were newly obtained with Nobeyama
45 m telescope). Locations of bright near-infrared (NIR) sources
identified by Kobayashi \& Tokunaga (2000) and MR-1 (Muzzio \& Rydgren
1974) are also indicated. In this infrared color image,
\textcolor{black}{
two clusters of red sources (Cloud 2-N cluster, Cloud 2-S cluster) were
identified in the two CO cores of the molecular cloud (Cloud 2-N and
Cloud 2-S).\footnote{These two CO cores were originally called Cloud
2a and Cloud 2b, respectively, in Digel et al. (1994).}  The Cloud 2-N
cluster was first discovered with the QUIRC image, while the Cloud 2-S
cluster was already recognized as an infrared source, IRS 2, in
Kobayashi \& Tokunaga (2000) and was resolved as a star cluster in the
QUIRC image.
}


We performed photometry of those red sources and found that many of them
have $H-K$ color excess in the $J-H$ vs. $H-K$ color-color diagram. We
present the photometric properties of those red sources in companion
papers by Yasui et al. (2006, 2008), in which we
performed photometry with much higher 
\textcolor{black}{photometric accuracy}
using deeper and higher-resolution $JHK$ images obtained with the Subaru
8.2 m telescope.

\subsection{Two Young Embedded Clusters}


Figure 2 shows the expanded images of both clusters.  The Cloud 2-N
cluster looks like a loose association similar to those in Taurus dark
cloud 
\textcolor{black}{(see, e.g., Lada, Strom, \& Myers 1993)}, 
while the Cloud 2-S cluster shows more packed morphology like those in
$\rho$ Oph star-forming regions (see, e.g., Allen et al. 2007),
suggesting the Cloud 2-N cluster is in {\it isolated star formation
mode}, while Cloud 2-S cluster is in {\it cluster star formation mode}.
The Cloud 2-N and Cloud 2-S clusters are distributed over a region of
$\sim$1.5 and $\sim$0.5 pc in diameter with stellar surface densities of
about 10 and 50 stars pc$^{-2}$, respectively, assuming the heliocentric
distance of 12 kpc. (See Yasui et al. 2008 for a more detailed analysis
based on deeper images.)  The density for Cloud 2-S is typical for young
embedded clusters, but that of Cloud 2-N is less than the lower boundary
for clusters listed in Lada \& Lada (2003).  There is a bright star at
the center of the Cloud 2-S cluster. Assuming a distance modulus of
15.4, the $K$-band luminosity of the brightest star is $M_\mathrm{K}$
$\sim$
\textcolor{black}{$-0.4$} 
%
, which is consistent with that of a late B-type
star (Tokunaga 2000).  Thus, the Cloud 2-S cluster seems to be a star
cluster aggregating around a relatively massive star, most likely a
Herbig Ae/Be star (see, e.g., Testi et al. 1999).
\textcolor{black}{
If the heliocentric distance is 16.6 kpc as derived from the recent
\ion{H}{1} data (Stil \& Irwin 2001), the above stellar densities are
reduced by a factor of $\sim$2, making the Cloud 2-N cluster a very
loose association ($\sim$5 stars pc$^{-2}$), while the Cloud 2-S cluster
is still a dense cluster with a typical density ($\sim$25 stars
pc$^{-2}$). Also, the absolute magnitude of the brightest star of the
Cloud 2-S cluster becomes $M_\mathrm{K} \sim -1.1$, which is consistent
with that of an early to mid B-type star (Tokunaga 2000). Therefore, the
above conclusion regarding isolated and cluster star formation modes
still holds despite the possible distance uncertainty.  }
%

\subsection{Isolated Bright Young Stellar Objects}
\label{subsec: isolated young stellar objects}


Kobayashi \& Tokunaga (2000) found several bright young stellar objects
(YSOs) in and around Cloud 2. Besides IRS 2 and 3, which are resolved as
clusters in Cloud 2-S, other IRS sources (IRS 1, 4, and 5) are also
detected and resolved in our new images (Fig. 3).  Those bright isolated
infrared stars are located eastward of the molecular cloud (see
Fig. 1). Their $K$-band absolute magnitudes ($M_\mathrm{K} = -2.4, -1.3,
-2.2$; Kobayashi \& Tokunaga 2000) and $H-K$ excess suggest all of them
are dust enshrouded intermediate-mass stars, probably Herbig Ae/Be
stars.  IRS 1, which is just outside of Cloud 2-N, appears to be a
single star, but it has an extended nebulosity in NE--SW direction (see
the left panel in Fig. 3), suggesting it is a YSO with outflow
activity. IRS 4 and 5, which are eastward of Cloud 2-S, have faint red
companion stars to the west (see the middle and right panels in Fig. 3),
suggesting they form typical small aggregations around intermediate-mass
stars (Testi et al. 1999).


\textcolor{black}{In our NIR images, IRS 1, 4, and 5, as well as MR-1, do
not show a cluster with a considerable number (e.g., $>$10) of NIR
sources at parsec scales as do the two embedded clusters. They appear to
be isolated bright objects.  Kobayashi \& Tokunaga (2000) argued that
those bright YSOs are associated with Cloud 2 because of their proximity
to the cloud and also because bright red sources were found only near
the Cloud 2 in their entire surveyed area of $34\arcmin \times
40\arcmin$. Also one of the sources, IRS 5, accompanies a small
molecular cloud core extending from the Cloud 2-S core toward the
southwest (Fig. 1). Therefore, their association with Cloud 2 appears to
be quite firm.
}

\section{STAR FORMATION IN CLOUD 2}


\subsection{Overall Activity and Geometry}
\label{subsec: Overall Activity and Geometry}


Besides the clusters and the isolated YSOs, other star
formation--related activities are known in the Cloud 2 region. Soon
after the discovery of Cloud 2, de Geus et al. (1993) discovered an
associated extended \ion{H}{2} region with an H$\alpha$ emission line,
and a visible
\textcolor{black}{early-type}
star, MR-1, which had been already found by an independent blue star
search in the outer Galaxy (Muzzio \& Rydgren 1974), was recognized as a
probable exciting source of the \ion{H}{2} region.  Stil \& Irwin (2001)
found extended 21 cm (1.4 GHz) continuum emission projected on Cloud 2-N
and that it is located on the maxima of the H$\alpha$ intensity map of
de Geus et al (1993). Based on the correlation with H$\alpha$ and the
absence of a counterpart at 74 cm, Stil \& Irwin (2001) suggested that
the continuum emission is thermal emission from an \ion{H}{2} region.


Figure \ref{fig: f4} summarizes the star formation activities in Cloud
2: the locations of NIR objects and MR-1 are overplotted on the
\ion{H}{1} 21 cm grayscale image,\footnote{From Canadian Galactic Plane
Survey (CGPS) data at
http://www1.cadc-ccda.hia-iha.nrc-cnrc.gc.ca/cgps/} 1.4 GHz radio
continuum coutour,\footnote{From NRAO/VLA Sky Survey (NVSS) data at
http://www.cv.nrao.edu/nvss/ (Condon et al. 1998).} and $^{12}$CO (1-0)
contour from our Nobeyama data. Figure 4 clearly shows that the CO
emission delineates the inner edge of the SNR \ion{H}{1} shell, whose
center is located toward the left bottom side of this image. Stil \&
Irwin (2001) first suggested that Cloud 2 is associated with the
approaching (blueshifted) side of the \ion{H}{1} shell based on the
coincidence of the sky position and the line-of-sight velocities
($v_\mathrm{LSR} = -103.6$ km s$^{-1}$ for $^{12}$CO and $= -101.9$ km
s$^{-1}$ for \ion{H}{1}).
\textcolor{black}{The CO cloud is located by the approaching side of the
\ion{H}{1} shell as if it delineates the inner side of the shell
(Fig. \ref{fig: f4}; see also Fig. \ref{fig: f5}).}
Because the radial velocities of CO and \ion{H}{1} clouds are consistent
with that for the \ion{H}{2} region ($v_\mathrm{LSR} = -101$ km s$^{-1}$
from Fabry-Perot H$\alpha$ imaging; de Geus et al. 1993), it is highly
likely that \ion{H}{1}, CO, and the \ion{H}{2} region are associated
with each other at the same distance from us. The radial velocity of the
visible
\textcolor{black}{early-type star} 
MR-1 was measured at 
\textcolor{black}{ $v_\mathrm{LSR} = -102.7 \pm 12$ km s$^{-1}$ by Russeil
et al. (2007), }
and this high velocity is also quite consistent with the velocities of
the warm and cold gas components. Some runaway OB stars with this
velocity have been observed in the I Per OB association (McLachlan \&
Nandy 1985), which is associated with the foreground Perseus arm but
relatively close to Cloud 2 on the sky with an angular distance of about
3--5 degrees. However, because most of the OB stars in the association
have $v_\mathrm{LSR} = -30$ to $-50$ km s$^{-1}$, which is the typical
velocity for the Perseus arm, it would be quite safe to say that MR-1,
with a much higher velocity, is associated with Cloud 2, which is
located far beyond the Perseus arm. We can therefore conclude that all
those objects in Figure \ref{fig: f4} (from \ion{H}{1} cloud to the NIR
clusters) are associated each other at the same distance.


De Geus et al. (1993) found that the three intensity maxima of the
H$\alpha$ (in their Fig. 3), which are shown with squares in Figure 4,
match well with the extended features of 1.4 GHz continuum, which
strongly support the suggestion by Stil \& Irwin (2001) that the radio
continuum is coincident with the \ion{H}{2} region in this star-forming
region.  The radio continuum emissions extend from near the CO cloud to
the east toward MR-1 (open star mark). This suggests that \ion{H}{2}
region is excited by MR-1 and partly by the IRS sources (filled green
star marks). We have also noticed that one more strong radio continuum
peak is located just 1--2 arcmin south of the Cloud 2-N cluster
(Fig. \ref{fig: f4}). This strong peak is not seen in the H$\alpha$
images by de Geus et al. (1993); rather, their image shows a local
minimum of the H$\alpha$ emission near this radio continuum
peak. Therefore, the \ion{H}{2} gas of this extended radio continuum
feature may be located {\it behind} the CO cloud, which should hide the
accompanying H$\alpha$ emission by strong dust extinction.  This view is
consistent with the fact that all the stars in the Cloud 2-N cluster are
behind the CO cloud with a uniform extinction of
\textcolor{black}{$A_V \sim 4$ mag as found by Yasui et
al. (2008):\footnote{\textcolor{black}{They originally suggested a total
extinction of $A_V \sim 7$ mag toward the Cloud 2-N cluster in Yasui et
al. (2006), but the value became slightly smaller ($A_V \sim 6$) after
re-analyzing the data (Yasui et al. 2008). The contribution from Cloud 2
only is $A_V \sim 4$ mag after subtracting the contribution from the
foreground ISM ($A_V \sim 2$ mag). This extinction is consistent with
Ruffle et al. (2007), who suggested $A_V < 4$ mag based on their
submillimeter observations and a lower than average dust-to-gas ratio in
the low-metallicity cloud.}}}
 we are seeing the Cloud 2-N cluster and the \ion{H}{2} emission feature
 through the moderately thick molecular cloud.  Unlike the Cloud 2-N,
 the Cloud 2-S does not seem to be under the strong influence of the
 photoionization by MR-1 and/or the bright NIR sources.


Figure \ref{fig: f5} shows Cloud 2 and the cloud members on the IRAS
60\,$\mu$m image\footnote{From IRAS Galaxy Atlas at IPAC, JPL
(http://irsa.ipac.caltech.edu/data/IGA/)} as well as the \ion{H}{1}
emission contour of the SNR shell. An extended IRAS source, IRAS
02450+5816, which is bright at 60 and 100\,$\mu$m, is located between
MR-1 and Cloud 2-N and was identified as a photodissociation region
formed by MR-1 and/or IRS 1 (Kobayashi \& Tokunaga 2000). The Cloud 2-S
cluster is in the vicinity of an unresolved IRAS source IRAS 02447+5811,
which is bright only in 60\,$\mu$m (see Table 2 in Kobayashi \& Tokunaga
2000) and is most likely related to the relatively hotter dust
around the bright YSO in the center of the Cloud 2-S cluster. There
might be some contribution to the IRAS flux of IRAS 02447+5811 from
other bright isolated YSOs, IRS 4 and 5, because its IRAS 60\,$\mu$m
image appears to extend slightly toward those bright isolated YSOs
(Fig. \ref{fig: f5}). We have checked mid-infrared images by the MSX
(Midcourse Space Experiment)\footnote{From MSX Data Atlas at IPAC, JPL
(http://irsa.ipac.caltech.edu/data/MSX/).} at 8--21$\mu$m, but could not
find any clear counterpart to the Cloud 2-N and 2-S clusters as well as
the bright isolated YSOs. Although the very extended 60\,$\mu$m features
at the top of Figure \ref{fig: f5} are accompanied mostly by
unrelated foreground clouds and star-forming regions, some of the
features (like the filamentary feature just outside the \ion{H}{1} shell
delineated by the dashed white line) might be related to the \ion{H}{1}
shell.

\subsection{Sequential Star Formation}
\label{subsec: Sequential Star Formation}


The CO cloud has an arc shape (Figs. 1, \ref{fig: f4}, and \ref{fig:
f5}) that is well aligned with the \ion{H}{1} SNR shell GSH 138-01-94
and open toward the center of the shell (Fig. \ref{fig: f5}), showing
that the CO cloud is clearly affected by the SNR \ion{H}{1} shell.  The
clusters are associated with the two densest CO peaks of Cloud 2, and
the sharp CO contours from the cluster toward the center of the SNR
shell (Fig. 1;
\textcolor{black}{see also CO 2-1 map in Ruffle et al. 2007)}
suggest that the compression of molecular gas triggered the star
formation. In Cloud 2-N, the cluster is associated with the northeastern
subpeak, which appears to be more strongly compressed in the expansion
direction than the southwestern subpeak, which does not harbor any clear
star formation activity.


\textcolor{black}{
The young stars in and around Cloud 2 appear to show an age sequence
from east to west following the direction of the SNR shell expansion.
Compared to the visible early-type star MR-1, the bright NIR sources IRS
1, 4, and 5 apparently form a younger stellar population with redder
colors and locations closer to the molecular cloud.  Although IRS 5 may
have a small molecular cloud core extending toward the southwest (see
Figs. 1 and \ref{fig: f4}), all three sources are not directly
associated with the massive molecular clouds as are the two embedded
clusters. While the clusters are visible only in the near-infrared, we
confirmed that IRS 1, 4, and 5 are marginally visible in the DSS2
$R$-band and IR-band plates. These facts suggest that the bright NIR
sources are less embedded and older than the embedded clusters.  In
summary, there is an age sequence (old to young) from the visible
early-type star, MR-1, to the isolated intermediate-mass infrared stars
IRS 1, 4, and 5, and finally to the two embedded clusters in the
molecular clouds. The direction of the age sequence following that of
the SNR shell expansion (see Fig. \ref{fig: f4}) suggests a {\it
sequential star formation} triggered by the supernova explosion.
}


\textcolor{black}{
Was the formation of MR-1 triggered by the SNR shell, or was it
coincidentally situated there before the SN explosion?  The age of the
SNR shell was originally estimated at 4.3 Myr by Stil \& Irwin
(2001). Because the expansion age is in proportion to the shell radius
in Stil \& Irwin's model,\footnote{\textcolor{black}{In their model the
expansion radius R is in proportion to t$^{\frac{2}{7}}$.}} it can be as
low as 3.0 Myr for our assumed distance ($D = 12$ kpc) or as much as 4.3
Myr for the Stil \& Irwin's assumed distance ($D = 16.6$ kpc). On the
other hand, it is difficult to estimate the age of MR-1. Its spectral
type was estimated at B0--B1 (Muzzio \& Rydgren 1974; Smartt et
al. 1996), suggesting that its mass is about 20 $M_\odot$ and its age is
no more than 10 Myr if it is on the main-sequence. Smartt et al. (1996)
suggested that judging from its low gravity, log $g = 3.7\pm0.1$, MR-1
is likely just evolving off the main sequence, while that of a B0--B1V
star is log $g = 3.9$ (Drilling \& Landolt 2000). In this case, the age
of MR-1 should be greater than 10 Myr, and the star should have been
located there before the supernovae explosion. However, Russeil et
al. (2007) recently re-estimated the spectral type of MR-1 as O9V (thus,
higher effective temperature with about 2000 K increase) with newly
obtained high-resolution spectra. This suggests that MR-1 could well be
still on the main-sequence. Because Smartt et al.'s suggestion was based
on a 0.2 dex difference in gravity (with the uncertainty of 0.1 dex) and
because there is the effective temperature uncertainty as shown by
Russeil et al.'s new observation, it would be hard to conclude that MR-1
is just evolving off the main-sequence. In this case, the age of MR-1
could be much less than 10 Myr.  Because there is no convincing method
to measure the age of a star on the main-sequence, it would be difficult
to precisely determine the age of MR-1 with the accuracy necessary to
compare its age with that of the SNR shell. Although there is a small
possibility that MR-1 formed in situ prior to the SN-explosion, we
suggest that the formation of MR-1 was triggered by the SNR shell in
view of the small probability that the sequence of young stellar objects
observed is by chance. In this scenario, the age of MR-1 can be as much
as the age of the SNR shell (3 Myr) and most likely about 2 Myr in view
of the location of MR-1 with respect to the current SNR shell (see
discussion in \S\ref{subsec: Star Formation} for the age estimate with
the SN-trigger model). In \S4 we are going to discuss the process of the
SN-triggered star formation in Cloud 2 in more detail, assuming MR-1 is
the first object formed by the SNR shell.}

\section{SUPERNOVA TRIGGERED STAR FORMATION}



\subsection{Cloud Formation}


Stil \& Irwin (2001) concluded that the expanding \ion{H}{1} shell, GSH
138-01-94, was formed by a supernova because of the nonexistence of any
OB association inside the shell. Although there is no conclusive
evidence for the supernova explosion itself, as there is for the nearby
SNR \ion{H}{1} shells with radio continuum and X-ray emission, this
interpretation appears to be quite solid in view of the observed
properties, e.g., the perfectly spherical shell with a very large
diameter that can survive only in a very low gas density environment
with little or no perturbation from other objects.


Stil \& Irwin (2001) suggested that Cloud 2 is associated with the
approaching (blueshifted) side of the \ion{H}{1} shell based on the
coincidence of the sky position (see Fig. \ref{fig: f5}) and the
line-of-sight velocities ($v_\mathrm{LSR} = -103.6$ km s$^{-1}$ for
$^{12}$CO and $= -101.9$ km s$^{-1}$ for \ion{H}{1}, while the center
velocity of the expanding shell is $v_\mathrm{LSR} = -94.2$ km s$^{-1}$
). Recently,
\textcolor{black}{Ruffle et al. (2007)} 
concluded that the SNR \ion{H}{1} shell has interacted with Cloud 2
based on extensively observed and modeled molecular chemistry for Cloud
2. \textcolor{black}{They} suggest that the chemistry of Cloud 2 is a
direct result of shock fronts from the SNR \ion{H}{1} shell propagating
through the cloud sometime between $10^3-10^4$ years ago (see
\textcolor{black}{also} Chapter 13 of Ruffle 2006). At the EOG, there is
also no confusion with other objects because there is no complex star
formation history and there is very low gas density and little
perturbation from the spiral arms. Therefore, we can conclude that the
\ion{H}{1} shell and the molecular cloud are closely associated.


\textcolor{black}{The shape of the molecular cloud follows the \ion{H}{1}
shell (Figs. \ref{fig: f4} and \ref{fig: f5}), and this suggests that
the formation of the molecular cloud itself was related to the SNR
shell.  Ruffle et al. (2007) considered the possibility that the cloud
was formed by the \ion{H}{1} shell from swept-up interstellar gas
through Rayleigh-Taylor instabilities. However, it is expected that the
formation of a molecular cloud is slowed down in the low-metallicity
environment because the formation of H$_2$ molecules requires dust
particles (e.g., see \S4.6 in Dyson \& Williams 1997), and there is less
dust in the low-metallicity environment. Typically, it takes 30 Myr to
form a giant molecular cloud even under conditions of solar metallicity
(e.g., Tielens 2005), and it could take even longer in Cloud 2 in view
of its very low dust content (Ruffle et al. 2007). This formation
timescale is much longer than the age of the SNR shell ($\sim$3 Myr).
Therefore, it is likely that the molecular cloud was already there
before the SN explosion and that the SN shock that passed through the
molecular cloud formed the shape of the cloud (P. Ruffle, private
communication), although we cannot completely exclude the possibility
that the molecular cloud was formed by the SN shock. The cloud formation
process itself in such low-metallicity environment still remains an
important open question.
}
%

\subsection{Star Formation}
\label{subsec: Star Formation}


The steep CO contour on the inner side of the SNR shell for both Cloud
2-N and 2-S (Figs. 1 and \ref{fig: f4}) strongly suggests that the
compression of the molecular cloud was brought on by the \ion{H}{1}
shell expansion. Some portion of the compression could be produced by
the stellar wind from the formation of the first star, MR-1, as a
secondary process of the supernova explosion.  The \ion{H}{2} region
(Fig. \ref{fig: f4}), which extends in and around Cloud 2-N, clearly
shows that at least Cloud 2-N is strongly affected by MR-1. However,
Cloud 2-S does not have either an associated H$\alpha$/radio continuum
(Fig. \ref{fig: f4}) or PDR (Fig. \ref{fig: f5}), which suggests that
Cloud 2-S was compressed purely by the SNR shell.


If the passage of the SNR shell has triggered the cluster formation in
Cloud 2, the upper-limit of the age of the cluster can be estimated from
the projected angular difference of the cluster from the current SNR
shell front. Figure \ref{fig: f6} shows that the projected angular
difference is about 300 arcsec. Because the radius of the \ion{H}{1} SNR
shell (2238 arcsec) and its age is estimated at \textcolor{black}{3.0} Myr
for our assumed distance $D = 12$ kpc (see discussion in \S\ref{subsec:
Sequential Star Formation}), the projected angular difference suggests
that the cluster is younger than 300/2238 $\times$
\textcolor{black}{$3.0 \sim 0.4$ Myr}.
Although the shell expansion speed is expected to decline with time,
that only shortens the upper-limit. Therefore, the age upper-limit of
\textcolor{black}{0.4}
Myr still holds. It is very interesting to note that this age estimate
is in quite good agreement with an independent age estimate of Cloud 2-N
cluster by Yasui et al. (2006), who estimated the age of the cluster at
about 0.5 Myr from the modeling of the $K$-band luminosity
function. This consistency strongly supports the idea that the cluster
formation was basically triggered by the SNR.
\textcolor{black}{Note that the SN shell would have moved more slowly
through the denser material of Cloud 2 because of mass conservation.
This possible slow expansion may cause some systematic uncertainty of
the above age estimate.}

\subsection{Isolated and Cluster Mode Star Formation}
\label{subsec: Isolated and Cluster Mode Star Formation}


The two embedded clusters show a distinct morphology difference: The one
in the northern molecular cloud core is a loose association with an
isolated star formation mode, while the other in the southern molecular
cloud core is a dense cluster with cluster star formation mode.  Because
the cloud mass of both Cloud 2-N and 2-S are similar ($\sim$5 $\times
10^3 M_\odot$;\footnote{Originally Digel et al. (1994) estimated the
cloud mass $\sim10^4 M_\odot$ assuming the kinematic distance of $D =
21$ kpc ($R_\mathrm{g} = 28$ kpc). The presented number was converted
from the original estimate assuming the distance of $D = 12$ kpc
($R_\mathrm{g} = 19$ kpc).
\textcolor{black}{ Ruffle et al. (2007) also calculate the cloud mass of
both Cloud 2-N and 2-S as $\sim$5$\times 10^3$ M$_\odot$ for the distance of
14 to 20 kpc ($R_\mathrm{g} =$ 22 to 28 kpc) using observed column
densities and LVG models.}
} Digel et al. 1994;
\textcolor{black}{Ruffle et al. 2007}),
the difference in the appearance of the two clusters may give us a good
opprotunity to study what causes the differences between the two star
formation modes.

In Figure \ref{fig: f7} we note that Cloud 2-N is distributed along the
tangential direction of the SNR shell, while Cloud 2-S extends slightly
inward. A close investigation of the SNR \ion{H}{1} shell in a velocity
range similar to that of the CO clouds ($-102$ to $-105$ km s$^{-1}$),
revealed that another shell-like structure is associated with Cloud 2-S
(Fig. \ref{fig: f8}).  This substructure in the \ion{H}{1} map can be
also seen in the \ion{H}{1} velocity channel map at $-104.5$ km s$^{-1}$
in Figure 1 in Stil \& Irwin (2001). We denote it as Cavity 2B since it
encompasses Cavity 2 in the Stil \& Irwin paper. A closer look at the
molecular cloud (Fig. \ref{fig: f7}) shows that the southern half of
Cloud 2 appears to be perturbed by Cavity 2B at around
(RA$_\mathrm{J2000.0}$,
Dec$_\mathrm{J2000.0}$)=(2$^\mathrm{h}$48$^\mathrm{m}$30$^\mathrm{s}$,
58$^\circ$26$'$).

We also found that both embedded clusters in Cloud 2 (Cloud 2-S cluster
and the IRS 3 mini-cluster) are located on the western side of the cloud
cores (see the distribution of red crosses in Fig. \ref{fig: f7}). This
appears to imply that the shock from the western side, presumably from
Cavity 2B, caused the star formation in Cloud 2-S. Therefore, we propose
that {\it strong compressions} of Cloud 2-S from the combination of the
SNR shell and Cavity 2B caused the cluster-mode star formation whereas
an isolated-mode star formation occurred in Cloud 2-N. This may be one
of the best examples that clearly supports the theoretical and
observational suggestions that high-pressure is the trigger of
cluster-mode star formation (e.g., Elmegreen 1998, 2004).

The center of Cavity 2B is at about (l, b) = ($137\fdg59$, $-1\fdg16$)
or (RA$_\mathrm{J2000.0}$, Dec$_\mathrm{J2000.0}$) =
(02$^\mathrm{h}$46$^\mathrm{m}$53$^\mathrm{s}$, +58$^\circ$23$'$) with a
radius of about 12$'$ (about 45 pc). Cavity 2B could be either a smaller
(thus, younger) adjacent SNR shell or a hole on the SNR shell like
Cavity 1 and 2 in Stil \& Irwin (2001). Although Cavity 1 appears to be
a hole on the SNR shell rather than another SNR shell (Stil \& Irwin
2001), we confirmed a sign of an expanding shell for Cavity 2B in an
\ion{H}{1} position-velocity map made with the CGPS data.
Because we could not identify any radio continuum source inside Cavity
2B using the NVSS map (VLA 1.4 GHz), Cavity 2B does not seem to be
powered by star-forming regions, and it could be another young SNR shell.
\textcolor{black}{Because this area around (l,b) = ($137\fdg5$,
$-1\fdg0$) on the SNR shell appears to be crowded with \ion{H}{1} clouds
throughout the velocity range of the expanding shell (see Fig. 1 in Stil
\& Irwin 2001), another interpretation is that Cavity 2B is just a group
of in situ \ion{H}{1} clouds that were distributed before the SN
explosion. Cloud 2 powered by the main SNR shell (GSH 138-01-94) may
have collided with the \ion{H}{1} clouds, which caused the strong
compression that stimulated the formation of the Cloud 2-S clusters.}

\subsection{Summary of the SNR-triggered Star Formation History }
\label{subsec: Summary of the SNR-triggered Star Formation History}


Figure \ref{fig: f9} summarizes our interpretation of the star formation
history in Cloud 2. The first-generation star (MR-1) and
second-generation stars (the bright NIR sources) do not seem to
accompany a cluster like that for the third-generation stars in Cloud
2-N and 2-S clusters. This may suggest that those early generation stars
were born in an isolated mode without accompanying clusters. Although
the ages of the clusters were estimated at $\sim$0.5 Myr (see discussion
in \S4.2), it is hard to determine when those early generation high- to
intermediate-mass stars were born. In view of their possible locations
inside the SNR sphere (the projected distance from the shell center to
the stars is about 75\% of the shell radius), their ages should not be
more than 2 Myr in view of the age of the SNR shell
(\textcolor{black}{$\sim$3 Myr; see \S4.2})
even after considering the slowing down of the shell expansion. This
probable age for the second generation bright NIR sources is consistent
with our interpretation that they are Herbig Ae/Be stars, whose typical
ages are 0.1 to a few Myrs (van den Ancker et al. 1997). The probable
age of $<$2 Myr is also consistent with the first generation star, MR-1,
which is most likely in the main-sequence phase in view of its very
short pre-main-sequence timescale ($\sim$0.1 Myr; Bernasconi \& Maeder
1996) for the large mass ($\sim$20 $M_\odot$).  Most of the red sources
scattered in the field in Figure 1 are background galaxies, although
some of them might be stars formed with the first and second generation
high/intermediate-mass stars. Note that the faint companions of the
bright NIR stars, IRS 4 and 5, are distributed to the west. This might
show the trace of the progressive star formation: first bright
intermediate-mass stars were born, then the low-mass stars just to the
west formed from the remaining cloud. Future kinematical study of all
those cloud members with NIR echelle spectroscopy may shed light on the
dynamical processes that happened during the triggered star formation.

\section{LINK TO THE EARLY PHASE OF THE GALAXY'S FORMATION}

\label{subsec: Link to the Early Phase of the Galaxy Formation}


Ferguson, Gallagher, \& Wyse (1998) have pointed out that the outskirts
of spiral galaxies in the local universe have the characteristics
similar to high-redshift damped Lyman-$\alpha$ systems as well as to
giant low surface brightness galaxies in that they have low gas surface
densities (yet high gas fractions compared to stars), low metallicities,
and long dynamical timescales. The irregular (Im) dwarf galaxies, which
are generally dominated in the optical by their younger stellar
populations, also have the similar characteristics (Hunter et
al. 2006). They are usually lower in luminosity and surface brightness,
more gas-rich with a lower metallicity and dust content, and form stars
without the benefit of spiral density waves. All these characteristics
suggest that the irregular dwarf galaxies are representative of the
nature of star formation in the early universe (Hunter et al. 2006;
Hunter \& Elmegreen 2004, 2006).  However, the distance to the nearest
Im galaxy (LMC) is 50 kpc, while the distance to the EOG is 10--20
kpc. The proximity of the EOG enables us to resolve star cluster members
with ground-based seeing resolution (see Yasui et al. 2006, 2008) and
spectroscopic study is also possible with the 8-meter class telescopes
even for low-mass stars (e.g., down to $K=19$ to 20 mag). The gas
content can be also studied in detail statistically (Snell et al. 2002)
and with various molecular lines (Lubowich et al. 2004; Ruffle et
al. 2005; Brand \& Wouterloot 2007). Therefore, the EOG is an excellent
place to study the star formation processes that were present during the
early epoch of galaxy formation and that are still present in Im
galaxies (Hunter et al. 2006).


The metallicity of outer galaxy disks, Im galaxies, low-metallicity blue
compact dwarf galaxies (Kunth \& Ostlin 2000), and damped Lyman-$\alpha$
systems are observed down to [Fe/H] $\sim -3$ (e.g., Pettini 2004). This
metallicity range traces the critical phases of early Galaxy formation
when the major components, such as halo, thick disk, bulge, and thin
disk, formed (e.g., Fig. 3 in Buser 2000; Freeman \& Bland-Hawthorn
2002). The typical metallicity of the outer galaxy regions ($-1.5 <$
[Fe/H] $< -0.5$) suggest that they represent the late phase of halo
formation and the early phase of thick disk formation.  Despite many
extensive studies (e.g., Bensby et al. 2007, and references therein),
the origin and role of the thick disk in our Galaxy are not well
understood even decades after its discovery (Gilmore \& Reid
1983). Although the study of abundance patterns recorded in very low
metallicity stars ([Fe/H] $< -2.5$) provides a vital clue to the star
formation process in the very early epoch of galaxy formation (Audouze
\& Silk 1995; Shigeyama \& Tsujimoto 1998; Tsujimoto \& Shigeyama 1998),
this successful ``archeological method'' cannot be applied to the
metallicity range of the thick disk because the abundance patterns of
the formed stars are significantly affected by the Galactic global
chemical evolution of the interstellar medium, from which the stars are
formed.  The observational study of the outer Galaxy region may directly
reveal the details of the star formation process during the formation of
the thick disk, which took place at $z \sim 2$ ($\sim$10 Gyr ago), to
shed light on the origin and role of this important Galactic component.


In the last decade, the formation of Population III stars (the ``first
stars'') has been extensively studied theoretically (Abel, Brian \&
Norman 2002; Bromm, Coppi, \& Larson 2002; Nakamura \& Umemura 2001) not
only for its own sake, but because of its direct relation to the
reionization of the early Universe and galaxy formation at an early
epoch (see, e.g., Yoshida 2006).  The recent advancement in the study of
the extreme metal-poor stars (Christlieb et al. 2002; Beers \&
Christlieb 2005), which are the potential relics of the Pop III stars,
boosted the examination of early-phase galaxy formation through the
study of abundance patterns recorded in very low metallicity stars at a
very early epoch. Although theoretical studies suggest that the star
formation physics in the metallicity range of the outer galaxy regions
should not significantly differ from that for the solar metallicity
(e.g., Omukai 2000), the observational study of the regions in this
``niche'' metallicity range establishes a good link between very low and
solar metallicity ranges, thereby revealing important astronomical
processes in this metallicity range.

%


It is most likely that SN-triggered star formation was the major process
by which stars formed in the early epoch of galaxy formation because of
the lack of other major star formation triggers such as density waves in
spiral galaxies.
\textcolor{black}{ Tsujimoto, Shigeyama \& Yoshii (1999) first formulated
the SN-triggered star formation in the early epoch of Galaxy formation
by utilizing the abundance pattern recorded in the very low metallicity
stars in our Galaxy (see also Tsujimoto, Shigeyama \& Yoshii 2002). They
successfully constrained the star formation efficiency and the mass
function during the formation of the Galactic halo.  With the clear
SN-triggered star formation signatures, Cloud 2 and probably other EOG
clouds are excellent places to confirm the predictions from near-field
cosmology.}

\section{CONCLUSION}
\label{sec: CONCLUSION}

\textcolor{black}{
We present a detailed study of star formation in Cloud 2, which is one
of the active star-forming regions in the EOG and is one of the farthest
star-forming regions, with a probable Galactic radius of $\sim$20 kpc.
Cloud 2 has a quite different environment from that of the solar
neighborhood, including lower metallicity, much lower gas density, and
small or no perturbation from spiral arms.  As such, it is a useful
analog for the star formation process in an early epoch of our Galaxy.
In particular, the study of the EOG may shed light on the origin and
role of the thick disk, whose metallicity range matches well with that
of the EOG.  
}

\textcolor{black}{
Our main results are
\begin{enumerate}
\item With new wide-field near-infrared (NIR) imaging that covers the
entire Cloud 2, we discovered two young embedded star clusters located
in the two dense cores of the cloud.
\item The two embedded clusters show a distinct morphology difference:
The one in the northern molecular cloud core is a loose association
with an isolated star formation mode, while the other in the southern
molecular cloud core is a dense cluster with a cluster star formation
mode.
\item Using our NIR and $^{12}$CO data as well as \ion{H}{1}, radio
continuum, and IRAS data from the archives, we show clear evidence of
sequential star formation triggered by the large nearby supernova
remnant, GSH 138-01-94.
\item We propose that the high compression resulting from a
combination of the SNR shell and an adjacent shell caused the dense
cluster formation in the southern core.
\end{enumerate}
}



\acknowledgments

We are grateful to UH 2.2 m telescope staff, especially John Dvorak and
Dr. Andrew Pickles for their support during the QUIRC observations.  We
also thank Takuji Tsujimoto for the fruitful discussions and suggestions
about star formation in the early epoch of the Galaxy's formation. 
\textcolor{black}{We thank the anonymous referee for careful reading and
thoughtful suggestions that significantly improved this paper.}
This
research was supported by a Grant-in-Aid for Encouragement of Scientists
(A) of the Ministry of Education, Science, Culture, and Sports in Japan
(No.11740128). NK was partly supported by a JSPS (Japan Society for the
Promotion of Science) postdoctoral fellowship for research abroad. NK is
also very grateful to the staff of the Institute for Astronomy,
University of Hawaii for their warm support during his stay in Honolulu.

\clearpage



\begin{figure}
\epsscale{1.0}
\plotone{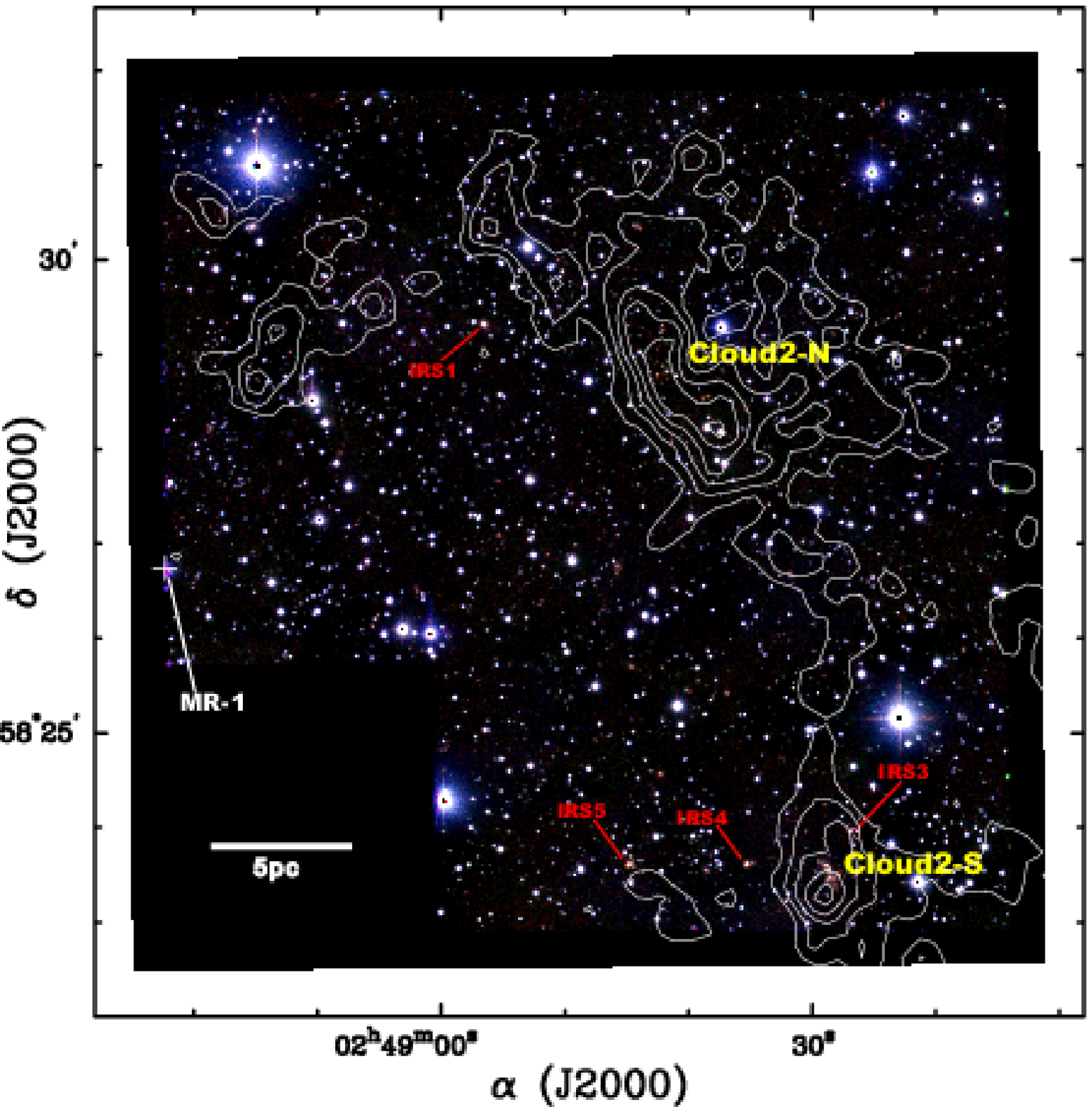}
\caption{
%
%
Near-infrared $JHK$ three-color image of Cloud 2 obtained with QUIRC
near-infrared camera. North is up, and east is to the left. $^{12}$CO
(1-0) data obtained with Nobeyama 45 m telescope is overplotted as the
white contour. The locations of bright NIR sources (IRS 1, 4, 5;
Kobayashi \& Tokunaga 2000) and a visible
\textcolor{black}{early-type}
star MR-1 are also indicated. A loosely packed embedded cluster (Cloud
2-N cluster) is seen as an aggregation of red sources in the northern CO
peak of the two CO peaks in the Cloud 2-N molecular cloud. A dense
embedded cluster (Cloud 2-S cluster) is seen near the CO peak of the
Cloud 2-S molecular cloud. This cluster was originally identified as IRS
2 in Kobayashi \& Tokunaga (2000). IRS 3 near Cloud 2-S was also found
to be a very small cluster (see also Fig. 2). Most of other faint red
sources in this field are faint background galaxies.  The NIR image was
Gaussian-convolved for viewing purpose. The resultant FWHM of the image
is about 0\farcs7.
\label{fig: f1}}
\end{figure}

\begin{figure}
\epsscale{0.75}
\plotone{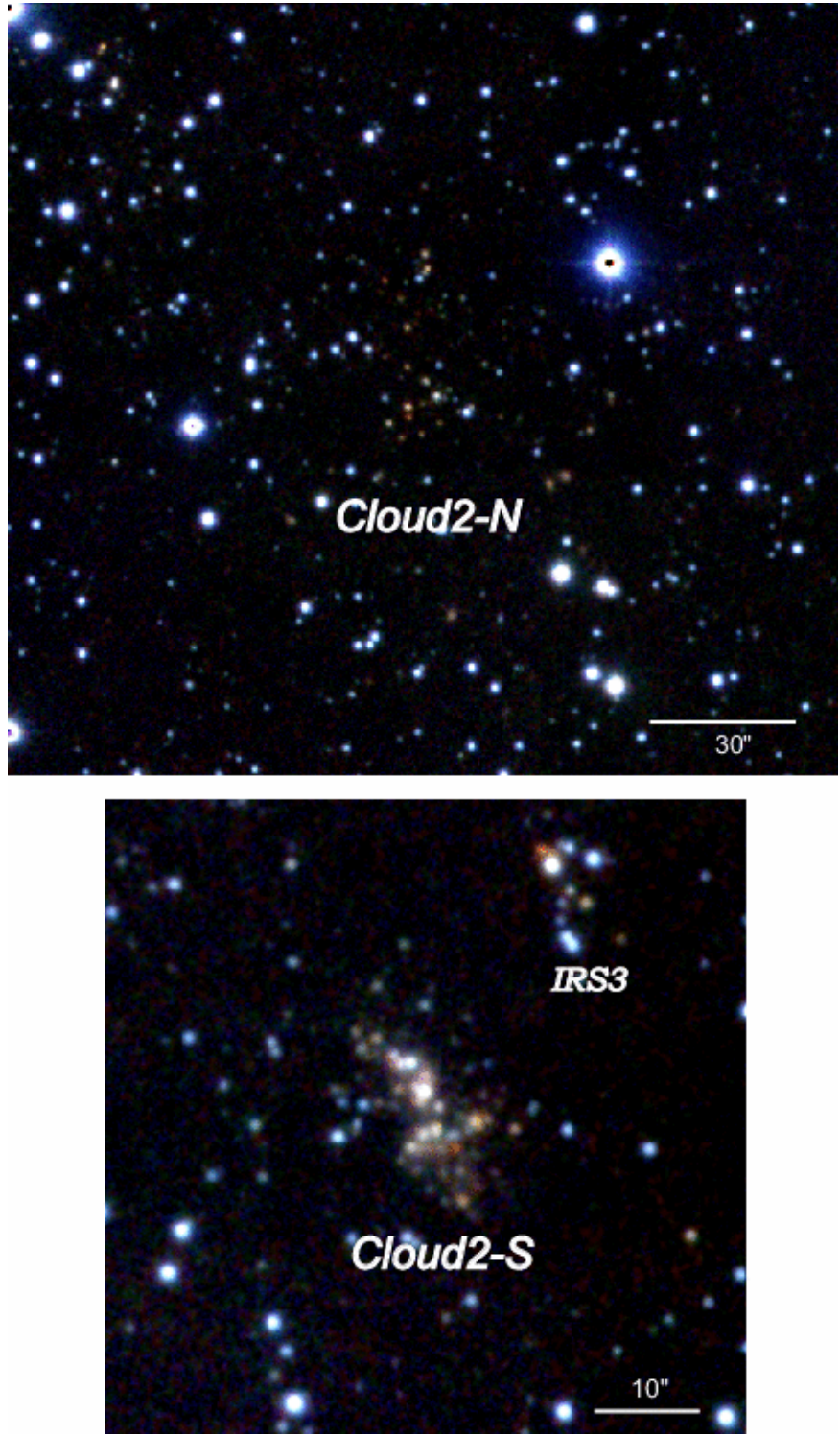}
\caption{
The expanded view of the $JHK$ three-color image of the two embedded
clusters ({\it Left}: Cloud 2-N cluster, {\it Right}: Cloud 2-S
cluster). Note the difference of stellar density in those two clusters.
IRS 3 near the Cloud 2-S cluster is resolved into a very small cluster
in this image.
\label{fig: f2}}
\end{figure}


\clearpage

\begin{figure}
\epsscale{1.0}
\plotone{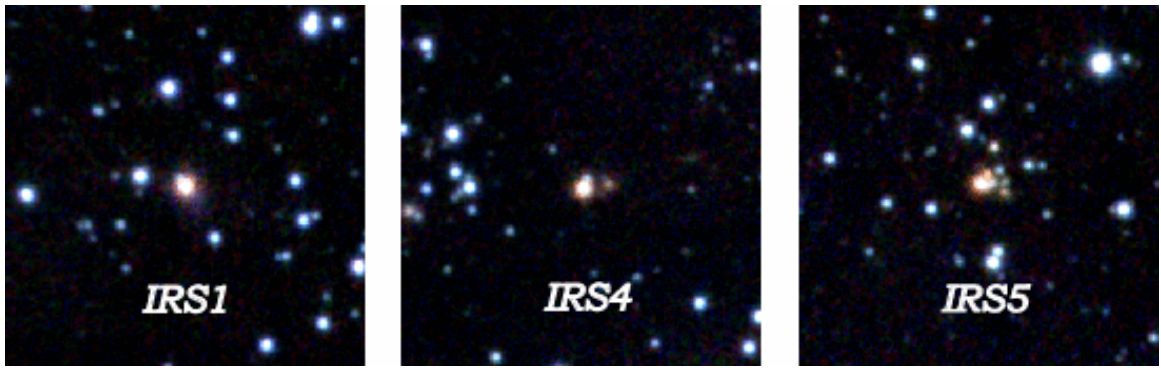}
\caption{
The expanded view of the $JHK$ three-color image of the bright NIR
sources near Cloud 2. Each box shows 30$''$ square. North is up, and
east is to the left.
These sources are identified as intermediate-mass YSOs (Herbig Ae/Be
stars). IRS 1 shows a faint nebulosity extended to southwest, which
would be an outflow or a reflection nebula. IRS 4 and 5 show several
faint red companion stars to the west, forming very small clusters.
\label{fig: f3}}
\end{figure}

\clearpage

\begin{figure}
\epsscale{1.00} 
\plotone{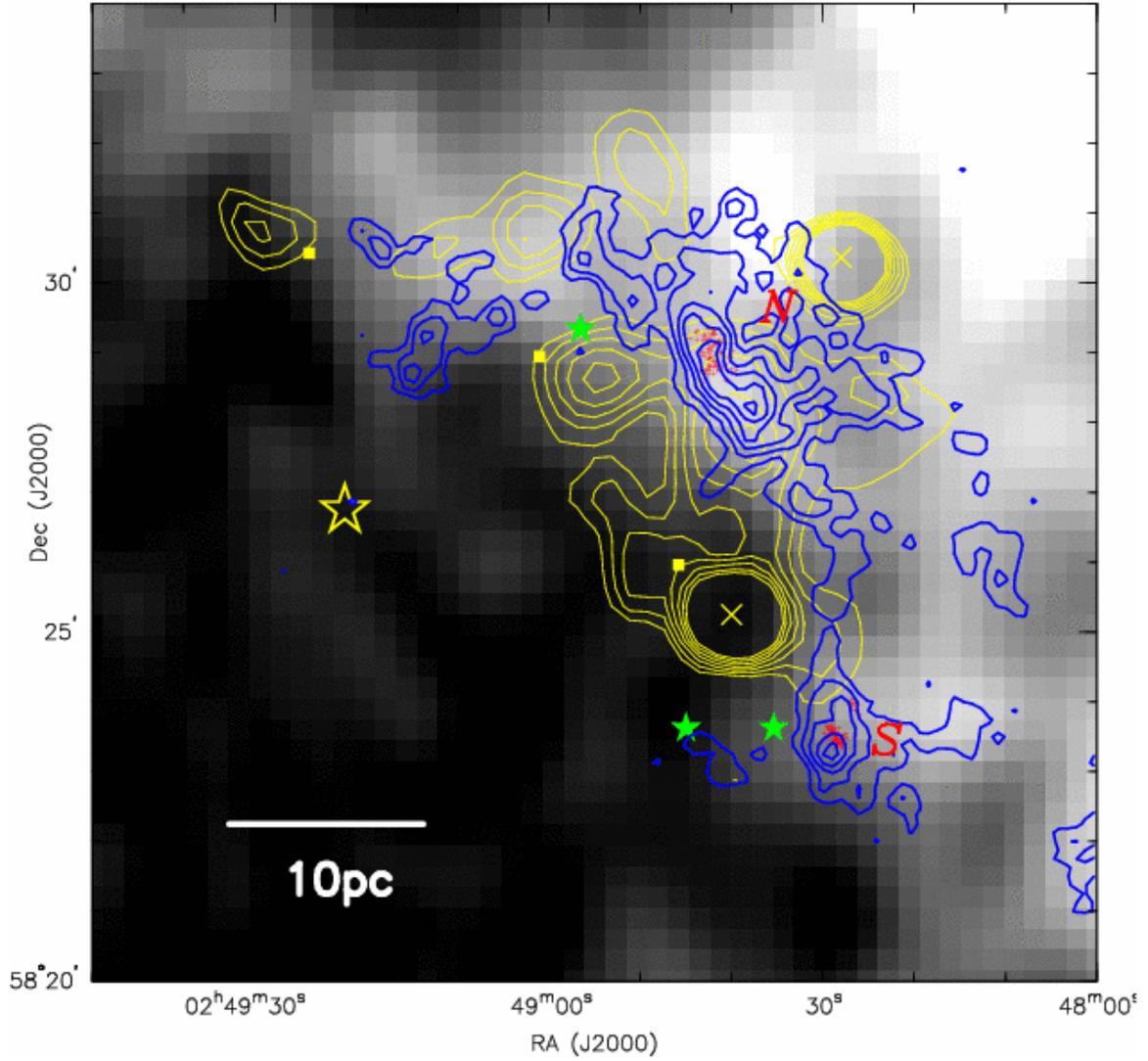}
\caption{
Summary of star formation activities in Cloud 2. A grayscale image and
blue contours show the distribution of \ion{H}{1} and $^{12}$CO (1-0)
emission integrated over the velocity range $-94$ to $-105$ km
s$^{-1}$. Yellow contours show 1.4 GHz radio continuum from NVSS (NRAO
VLA Sky Survey) data: Note that the two brightest pointlike sources with
cross marks are unrelated foreground/background radio sources. Filled
yellow squares show the peaks of H$\alpha$ nebula (de Geus et
al. 1993). The open yellow star and filled green stars show MR-1 and
bright NIR sources, respectively. The locations of the members in the
Cloud 2-N, Cloud 2-S clusters, and the Cloud 2-S mini-cluster are shown
with red crosses.
\textcolor{black}{[Refer to the ApJ on-line PDF to see the detail of the
distribution of the cluster members]}
 Although the spatial-resolution of the \ion{H}{1} image and the VLA map
 are originally 1$'$ and 45$''$, respectively, here they are
 Gaussian-convolved with 1-pixel (18$''$ and 15$''$, respectively) for
 better viewing. The beamsizes of the Nobeyama CO map is about 20$''$.
 See the discussion in the main text for detail.
\label{fig: f4}}
\end{figure}

\clearpage

\begin{figure}
\epsscale{0.75} 
\plotone{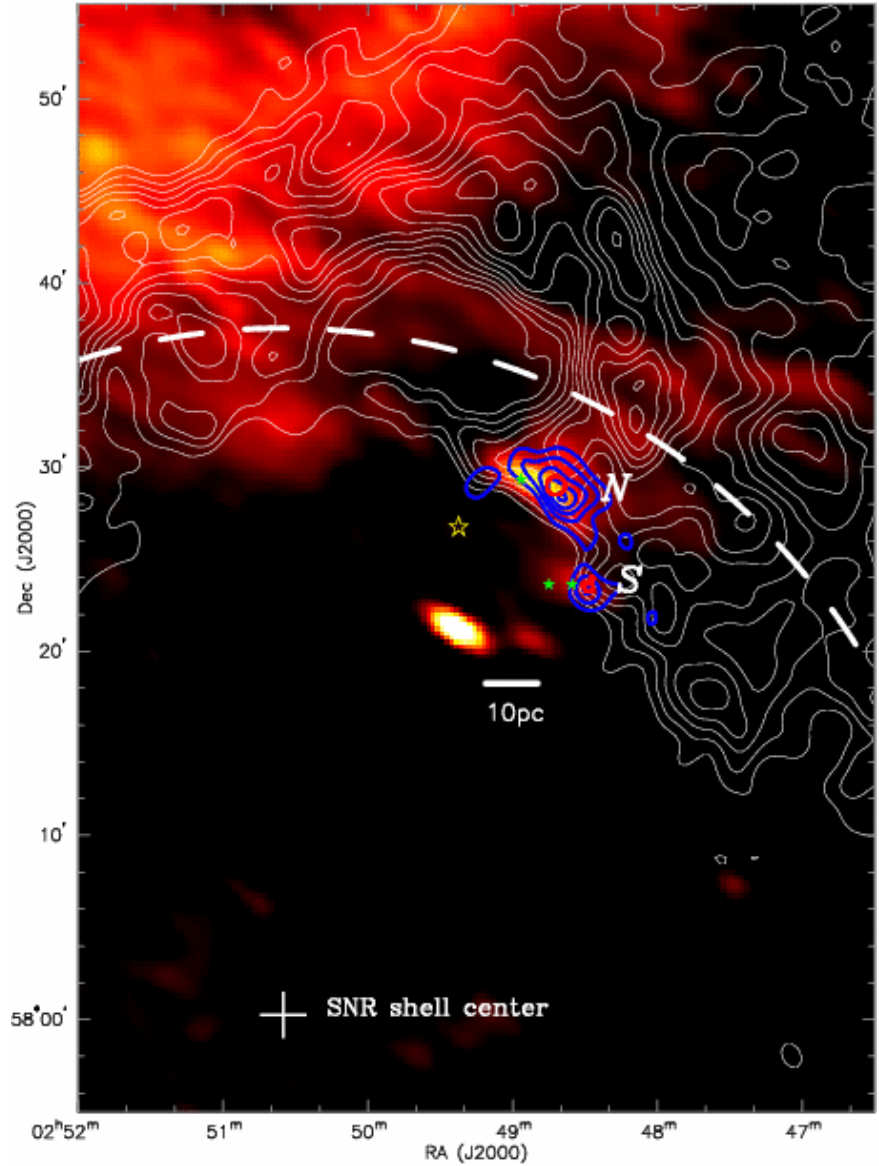} 
\caption{
The locations of Cloud 2 and forming stars with respect to the SNR shell
GSH 138-01-94 (Stil \& Irwin 2001). The center of the shell is shown
with a white cross, while the edge of the shell is shown with a dashed
line.  White contours show the \ion{H}{1} surface brightness integrated
over the velocity range $v_\mathrm{LSR} = -94$ to $-105$ km s$^{-1}$
(from CGPS data), while the blue contours show the $^{12}$CO (1-0)
surface brightness (from our Nobeyama data). The background image is the
IRAS 60\,$\mu$m image from IRAS IGA data.  The open yellow star mark,
filled green star marks, and red circles show the locations of the
visible
\textcolor{black}{early-type}
star MR-1, the bright NIR sources, and the two embedded clusters,
respectively.
The \ion{H}{1} and IRAS images are 1- and 2-pixel Gaussian-convolved,
respectively, for better comparison to the CO map.
The bright IRAS source in the center of the image is an unrelated
foreground object (a combination of two bright NIR sources, IRS 6 and 7
in Kobayashi \& Tokunaga 2000), and shows the average PSF of this IRAS
image.
See the discussion in the main text for detail.
\label{fig: f5}}
\end{figure}
\clearpage

\begin{figure}
\epsscale{1.00} 
\plotone{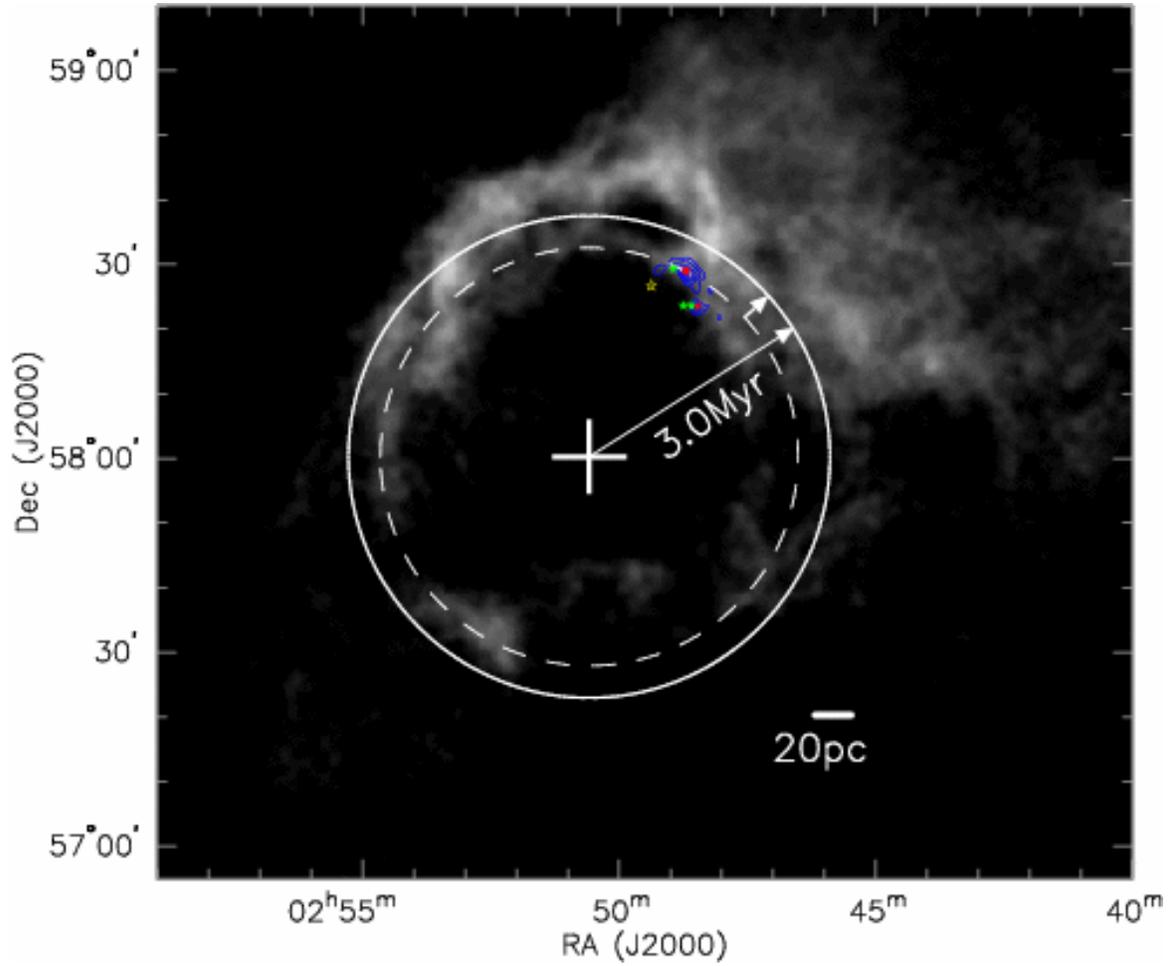} 
\caption{
The location of Cloud 2 and the star clusters in the SNR shell GSH
138-01-94. The grayscale image and blue contours show the \ion{H}{1} and
CO in $v_\mathrm{LSR} = -94$ to $-105$ km s$^{-1}$. The filled red
circles show the location of the young embedded clusters. The solid
circle shows the current location of the SNR shell with an
estimated age of \textcolor{black}{3.0 Myr (see \S4.2)}.  The dashed
circle shows the location of the Cloud 2-N cluster. The small arrow
shows the projected separation of 300 arcsec, from which the lifetime of
the cluster can be estimated as 0.4 Myr (see \S4.2 for detail).
\ion{H}{1} and CO data are same as that for Fig. \ref{fig: f5}.
\label{fig: f6}}
\end{figure}
\clearpage

\begin{figure}
\epsscale{1.00} 
\plotone{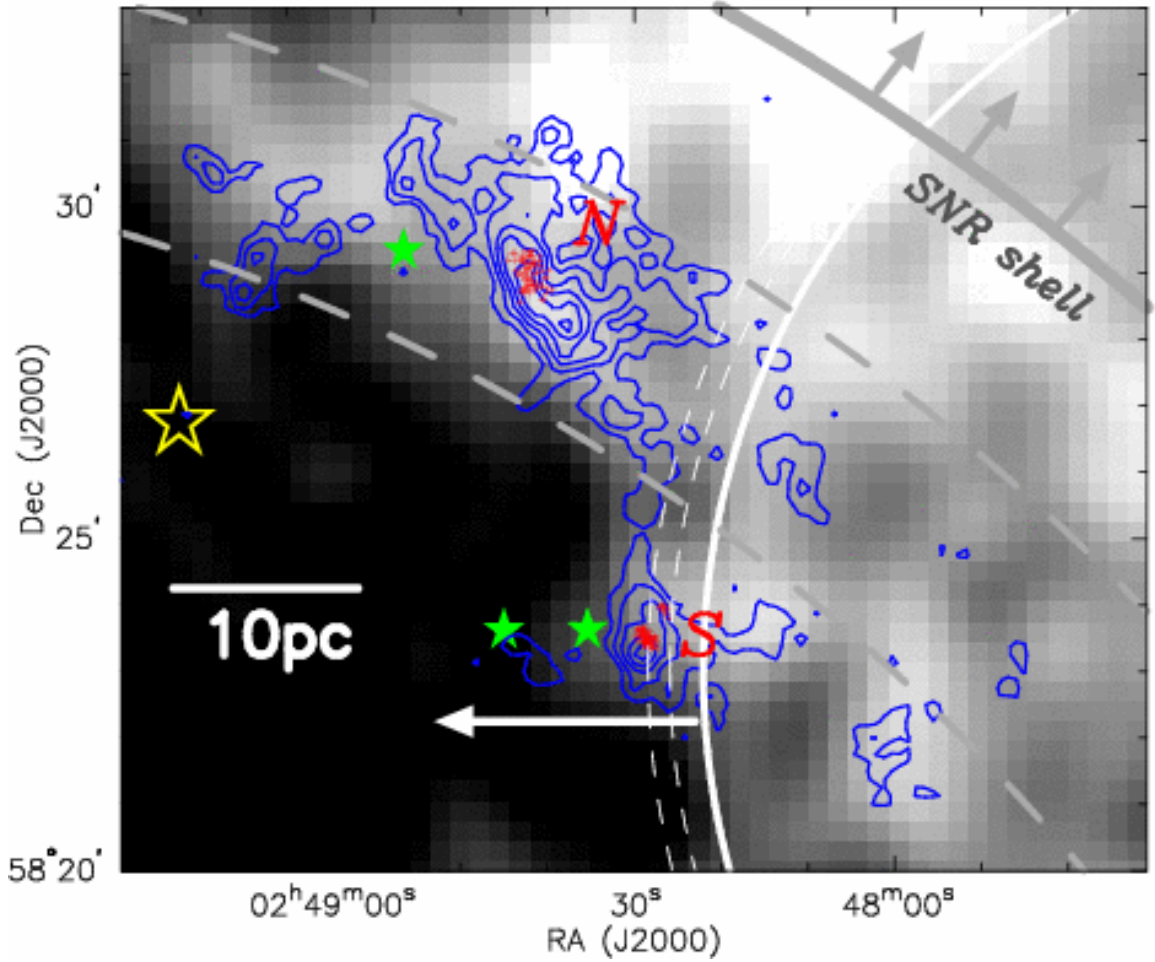} 
\caption{
The relationship of the \ion{H}{1} shells, Cloud 2 (blue contours), and
the embedded clusters (red crosses).  The outer edge of the SNR shell,
GHS 138-01-94, is a thick gray curve, while the edge of the Cavity 2B
(see the main text) is a thick white curve. Cloud 2-N is distributed
between the two dashed gray curves whereas Cloud 2-S appears to be
pushed back to inside the shell by the expansion of the Cavity 2B. The
cluster-mode star formation in Cloud 2-S occurs at the shock front
facing Cavity 2B. The \ion{H}{1} and CO images are the same as for
Fig. \ref{fig: f4}.
\textcolor{black}{[Refer to the ApJ online PDF to see the details of
 the distribution of the cluster members]}
The open yellow star mark and filled green star marks show the locations
of the visible early-type star MR-1 and the bright NIR sources,
respectively.
\label{fig: f7}}
\end{figure}
\clearpage

\begin{figure}
\epsscale{1.00} 
\plotone{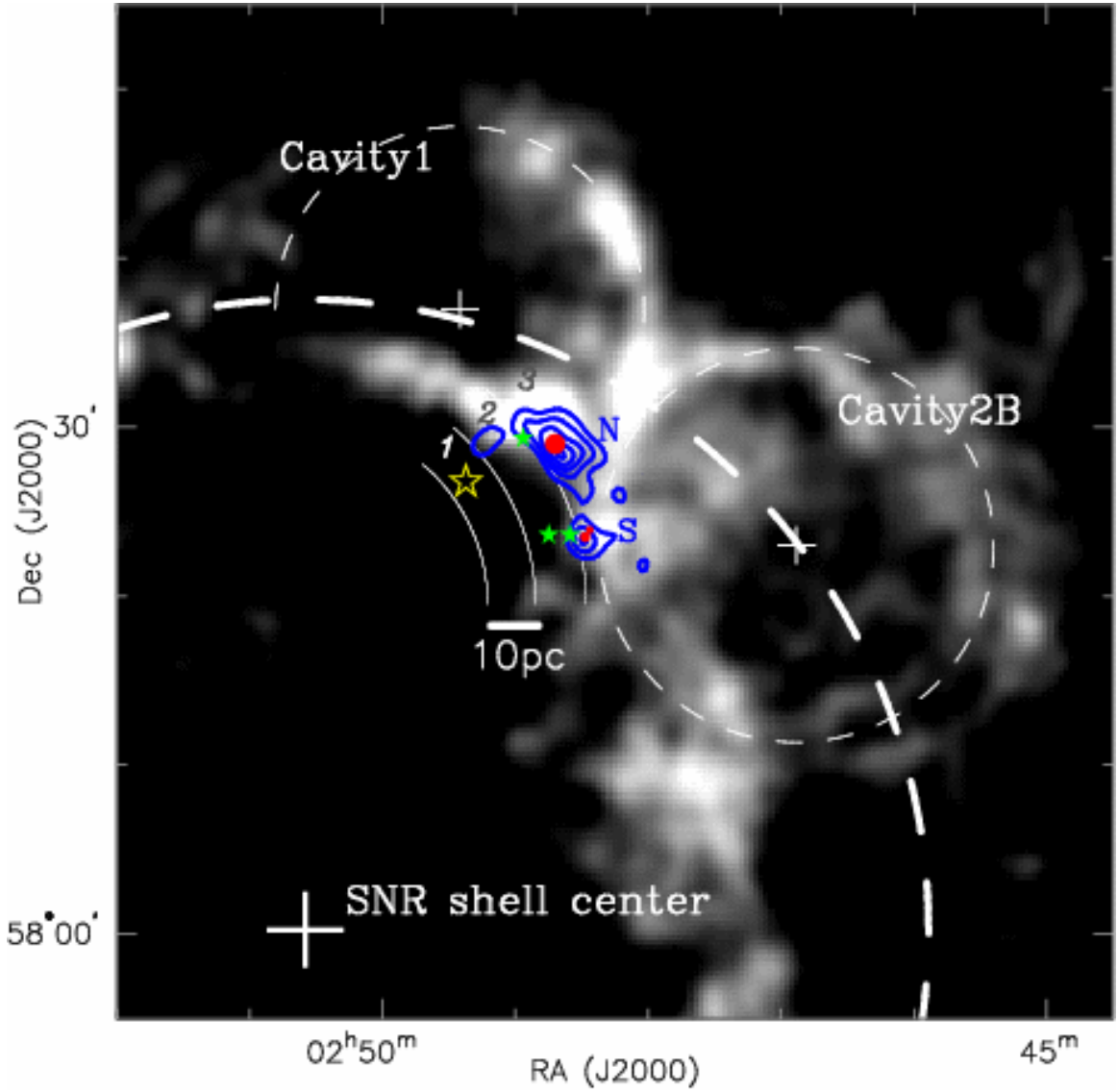} 
\caption{
Cloud 2 and the associated members are shown in the \ion{H}{1} 21 cm
grayscale image. Cloud 2 is shown with $^{12}$CO contours and its
members are shown with a yellow open star (visible
\textcolor{black}{early-type}
star MR-1), filled green stars (bright NIR sources), and red circles
(two embedded clusters). Progressive star formation from the inner side
of the shell toward the edge (from zone 1 to 2, then 3) is visible in
this image.
CO data is same as that for Figs. 1 and 4 except for the contour
interval. The \ion{H}{1} data from CGPS is integrated over a velocity
range from $-102$ to $-105$ km s$^{-1}$, which shows the adjacent
cavities clearly.
\label{fig: f8}}
\end{figure}
\clearpage

\begin{figure}
\epsscale{1.00} 
\plotone{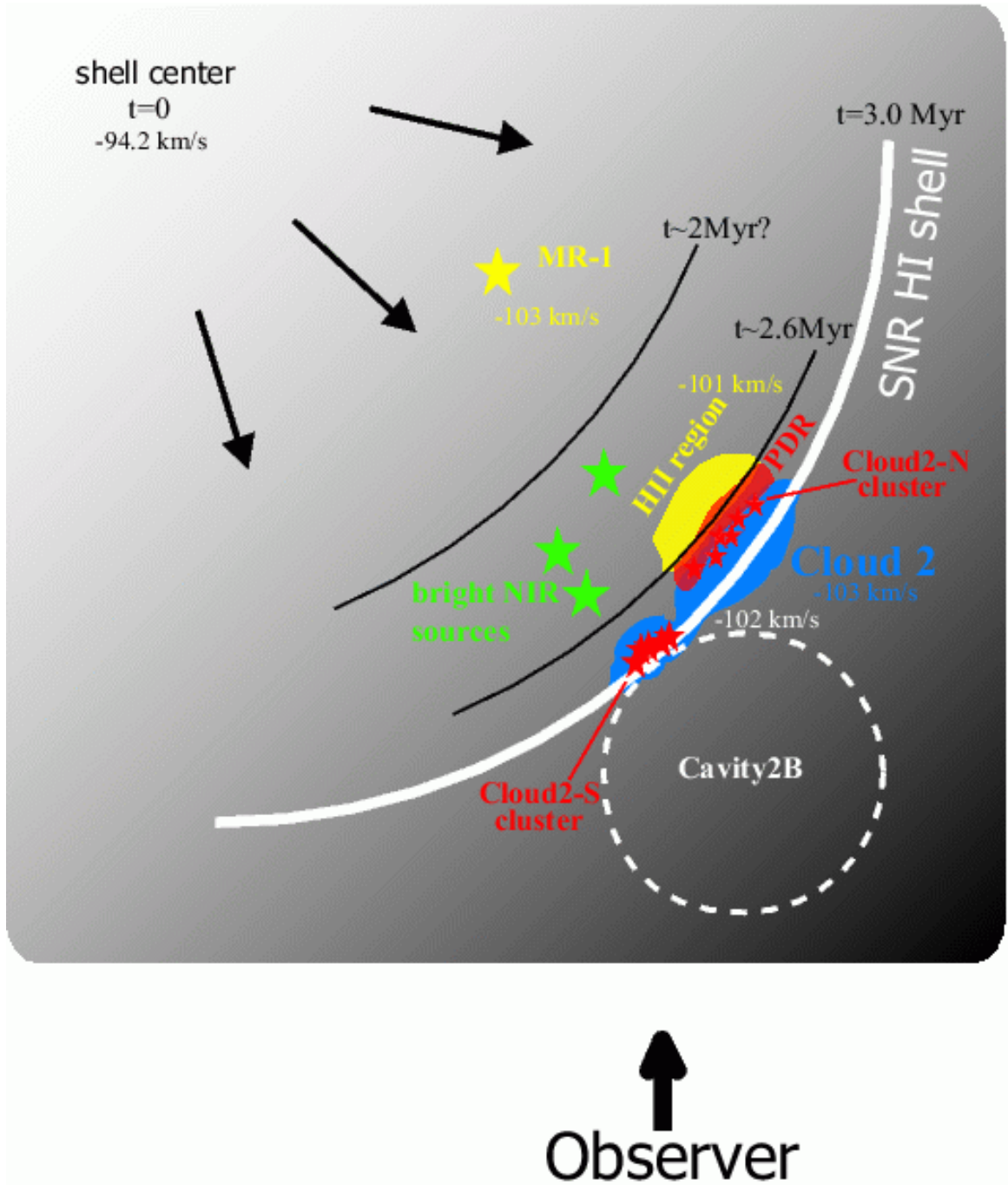} 
\caption{
A schematic view of the progressive star formation by the SNR shell. The
scales and the viewing angles are arbitrary.  The times shown are
counted from the time of the explosion of the SN. The radial velocities
of the objects are shown (shell center, \ion{H}{1}, and Cloud 2, Stil \&
Irwin 2001; \ion{H}{2}, de Geus et al. 1993; MR-1, Russeil et
al. 2007). See the main text for the detailed discussion.
\label{fig: f9}}
\end{figure}
\clearpage

\end{document}